\documentclass[reprint, superscriptaddress]{revtex4-1}

\usepackage[utf8]{inputenc}
\usepackage{t1enc} 
\usepackage{graphicx}   % need for figures
\usepackage{amsmath}
\usepackage{multirow}
\usepackage{xcolor}

\begin{document}

\title{Spin-polarized currents driven by spin-dependent surface screening}
\author{Piotr Graczyk}
\email{graczyk@amu.edu.pl}
\affiliation{Institute of Molecular Physics, Polish Academy of Sciences, M. Smoluchowskiego 17, 60-179 Poznan, Poland}
\author{Maciej Krawczyk}
\email{krawczyk@amu.edu.pl}
\affiliation{Faculty of Physics, Adam Mickiewicz University in Poznan, Umultowska 85, 61-614 Poznan, Poland}

\begin{abstract}

We have examined spin polarization of the electron current in a ferromagnetic metal induced by the spin-dependent surface screening at the dielectric-ferromagnetic metal (D-FM) interface. In an applied ac voltage, the dynamic band splitting driven by the changes in the screening charge at the D-FM interface develops spin accumulation. The resultant spin accumulation gradient produces a time-dependent spin current. We have derived the rate of the spin accumulation dependent on the rate of the screening charge density accumulation within the Stoner band model. The spin-charge dynamics in the system is then modeled by a set of diffusive equations with the contributions from spin-dependent surface screening and spin-dependent conductivity. We show for MgO-Cu-Co-MgO system that the spin-dependent screening in thin Co film produces spin accumulation 7 times higher than that obtained by the spin-dependent conductivity in thick Co films. We propose an experimental approach to validate our numerical predictions and to distinguish between spin accumulation induced by the spin-dependent conductivity and by the spin-dependent surface screening.

\end{abstract}

\maketitle

The phenomenon of spin-dependent surface screening (SS) leads to the charge-mediated magnetoelectric effect (ME) which is the key mechanism for the electric control of magnetic properties in ultrathin multiferroic heterostructures \cite{Burton2012,Vaz2012}. The effect has been described first theoretically by Zhang in 1999 \cite{Zhang1999} and then confirmed by \textit{ab initio} calculations \cite{Duan2008,Niranjan2009,Rondinelli2008}. It has been shown experimentally that SS may be utilized to change the magnetization, anisotropy or coercivity of ultrathin ferromagnetic film \cite{Molegraaf2009,Weisheit2007,Maruyama2009,Endo2010,Shiota2009,Shiota2011,Seki2011,Chiba2011}, to induce magnetic ordering transition \cite{Burton2009,Vaz2010} or even induce surface magnetism in nonmagnetic materials \cite{Sun2010}. In multiferroic tunnel junctions SS provides significant contribution to the tunneling resistance \cite{Zhuravlev2010,Yin2015} and opens possibilities to manipulate the spin polarization of tunneling electrons \cite{Pantel2012,Sen2019}. Owing to the surface magnetic anisotropy that is controlled by SS, spin waves may be excited parametrically or resonantly with the ac voltage \cite{Rana2017, Verba2014, Verba2017, Chen2017, Nozaki2012}.

Spin-dependent surface screening changes the absolute value of magnetization at the dielectric-ferromagnetic metal (D-FM) interface, in contrast to, studied more extensively, strain-driven ME effects \cite{Vaz2012,Graczyk2015,Graczyk2016,Graczyk2016a}, which rotate the magnetization direction. Although strain-driven ME remains significant for relatively thick layers, it suffers from ageing effects, clamping and integration problems with current technology. 

In this letter we study the generation of spin-polarized currents by spin-dependent surface screening effect in D-FM-D system subjected to the ac voltage. The change of the voltage results in the change of the screening charges at the insulator-ferromagnet interface. Since the screening is spin-dependent, this leads to the appearance of nonequilibrium spin density at the interface and then formed spin-density gradient generates diffusive spin current. We discuss this effect within the Stoner band model and derive the term that describes SS in drift-diffusive formalism. With this model we perform numerical simulations and find the spin current with the contributions from spin-dependent conductivity (SC) and spin-dependent surface screening (SS) for MgO-Cu-Co-MgO system. Finally, we propose the strategy to distinguish experimentally between that two current polarization mechanisms.

\begin{figure}
\includegraphics*{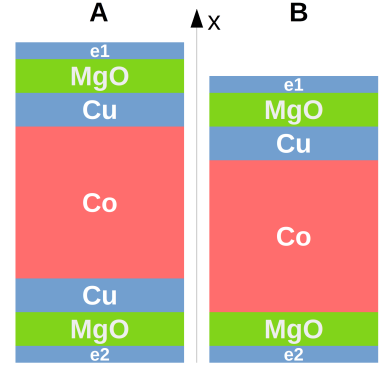}
 \caption{\label{Fig1} The schematic of considered multilayers. In system A there is no D-FM interface while in system B the single D-FM interface is present.}
\end{figure}

Considered systems are shown in Fig. 1. The multilayer consists of two 2-nm-thick MgO films separated by Co film of different thicknesses. In system A the Co layer is separated from MgO by 2-nm-thick nonmagnetic metal (Cu) at both interfaces while in the system B it is separated from MgO only from the top. Therefore, in the system A there is no D-FM interface, while in  the system B it is present. Consequently, screening charges will appear in the ferromagnet only at the bottom interface in the system B when the voltage is applied to the multilayer structure through bottom and top electrodes (e2 and e1, respectively). It is assumed that charge and spin transport occur only perpendicular to the layers, along $x$ axis.

The coupled charge-spin dynamics driven by the prescribed ac voltage is considered on the base of the diffusive model \cite{Valet1993,Zhu2008}. The free charge density current is described in the metal layers by the equation:
\begin{equation}
J_f=\sigma E - D \frac{\partial n_f}{\partial x}+\beta D \frac{e}{\mu_B} \frac{\partial s}{\partial x} ,
\label{eq1}
\end{equation}
which includes contributions from electron drift driven by electric field $E$, diffusion driven by the gradient of free charge density $n_f$ and diffusive spin polarisation (SC), respectively. Here, $\sigma$ is conductivity, $D$ is diffusion constant, $\beta$ is spin asymmetry coefficient, $\mu_B$ is Bohr magneton and $e$ is charge of electron. The displacement current is described by the equation:
\begin{equation}
J_b=\epsilon_0 \epsilon_r \frac{\partial E}{\partial t},
\label{eq1}
\end{equation}
where $\epsilon_0\epsilon_r$ is permittivity of the material.  The conservation of free ($i=f$) and bound ($i=b$) charge density $n_i$ is described by continuity equation:
\begin{equation}
\frac{\partial n_i}{\partial t}=-\frac{\partial J_i}{\partial x}.
\label{eq2}
\end{equation}

Electric potential $V$ is given by the Gauss Law: 
\begin{equation}
\begin{split}
\epsilon_0\frac{\partial^2 V}{\partial x^2}&=n_f+n_b, \\
E&=-\frac{\partial V}{\partial x},
\label{eq3}
\end{split}
\end{equation}
with the boundary conditions at the electrodes:
\begin{equation}
\begin{split}
V|_{x=e1} &=U=U_0 \sin (2 \pi f_0 t), \\
V|_{x=e2} &=0,
\label{eq4}
\end{split}
\end{equation}
where $f_0$ is the frequency of the oscillating voltage $U$.

The spin current $J_s$ in the ferromagnetic material is modeled by the equation \cite{Zhu2008}:
\begin{equation}
J_s=- D \frac{\partial s}{\partial x}-\beta\frac{\mu_B}{e}\left(\sigma E - D \frac{\partial n_f}{\partial x}\right) ,
\label{eq5}
\end{equation}
which describes spin current driven by gradient of spin accumulation $s$ (diffusion) and spin-conductivity term.

The continuity equation for the spin accumulation $s$ is:
\begin{equation}
\frac{\partial s}{\partial t}=-\frac{\partial J_s}{\partial x}-\frac{s}{T_1}+f_s,
\label{eq6}
\end{equation}
which describes the rate of change of $s$ due to gradient of the spin current, the spin-flip relaxation with the characteristic time $T_1$. $f_s$  is an additional source from spin-dependent charge screening. The formula for $f_s$ is derived below. We use in the following paragraphs $n \equiv n_f$.

We will consider the voltage-driven magnetization and spin density change in cobalt within the Stoner model \cite{Stoner1938}. Due to the exchange splitting of the band \textit{d} for the majority and minority spins and partially filled minority subband (Fig. 2a) there exists equilibrium spin density (magnetization) in the bulk of the ferromagnet which is equal to the difference between charge densities of the spin-up ($\uparrow$) and spin-down ($\downarrow$) subbands, $m=\mu_B(n^\uparrow-n^\downarrow)/e$. The voltage applied to the system shifts the potential of the bands at the D-FM interfaces (Fig. 2b) leading to the accumulation of the screening charge density $n=\delta n^\uparrow_0+\delta n^\downarrow_0$ where $\delta n_0^\sigma$ is accumulated screening charge in the subband $\sigma$. Since the density of states are different between subbands at the Fermi level ($\Delta \rho=\rho^\uparrow-\rho^\downarrow <0$ for Co), then screening charge changes the value of the magnetization at the interface by  $\delta m_0=\mu_B(\delta n^\uparrow_0-\delta n^\downarrow_0)/e$ \cite{Duan2008}.

\begin{figure}
\includegraphics*{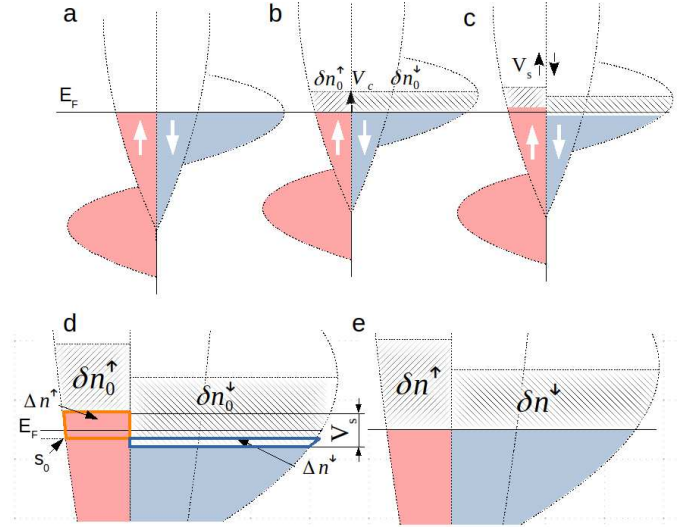}
 \caption{\label{Fig2} The schematic diagrams of \textit{s} and \textit{d} bands in Co (a) at electric field $E=0$ and in equilibrium; (b) at $E\neq 0$ with equal band shift $-V_c$; (c) at $E\neq 0$ with screening exchange splitting $V_s$. (d) Dynamic nonequilibrium charges (with respect to equilibrium spin level $s_0$)  $\Delta n^\uparrow$ and $\Delta n^\downarrow$ developing as a consequence of the dynamic exchange splitting. In (e) the equilibrium state with the exchange splitting of the screening charges is shown.}
\end{figure}

Another important feature of the screening in the ferromagnet is that the changes in the band occupancy influence the exchange splitting \cite{Niranjan2009}, i.e. the potential is spin-dependent \cite{Zhang1999,Zhuravlev2010}:
\begin{equation}
V^\sigma=V_c-\frac{I}{e}(\delta n^\sigma - \delta n^{-\sigma}),
\label{eq7}
\end{equation}
where $I$ is exchange constant. Spin-dependent potential (8) means that the subbands shift relative to each other by $V_s=V^\uparrow-V^\downarrow=-2I \delta m$.  (Fig. 2c). From Eqs. (5) and (11) in Ref. \cite{Zhang1999} we get the relation between $V_s$ and screening charge  $n=\delta n^\uparrow+\delta n^\downarrow$ in the linearized Thomas-Fermi model:
\begin{equation}
V_s=-\frac{2I\Delta \rho}{\rho+4 I \rho^\uparrow \rho^\downarrow}n,
\label{eq8}
\end{equation}
where $\rho=\rho^\uparrow+\rho^\downarrow$. This shift has important consequences. Since $\rho^\uparrow \neq \rho^\downarrow$, the shift changes the equilibrium charge level with respect to the Fermi level by $E_F-s_0$ (Fig. 2d) where $s_0=-e(V^\uparrow\rho^\uparrow+V^\downarrow\rho^\downarrow)/\rho$ is spin equilibrium level. This leads to the changes in equilibrium charge densities $\delta n^\sigma \neq \delta n_0^\sigma$ and potential $V_c$. Therefore, the surface magnetization value changes by $\delta m \neq \delta m_0$.  Moreover, the screening length in ferromagnet $\lambda_F$ deviates from the conventional Thomas-Fermi value $\lambda=\left(\epsilon_0/e^2 \rho \right)^{1/2}$ and thus the capacitance is reduced \cite{Zhang1999}. We can implement this effect to our numerical model where the screening length is defined as $\lambda=\left(\epsilon_0 D/\sigma \right)^{1/2}$ by dividing drift term and multiplying diffusion term in Eq. (1) by $\gamma_n=\lambda_F/\lambda \approx 1.1$ for Co. We found that modification of the screening length does not influence the effect of the spin-polarized current generation which is considered here so it will be neglected in further part.

Since screening charge and thus $V_s$ (Eq. (9)) is time-dependent under the ac volatage, the change of the shift between subbands $\Delta V_s$ is the source of nonequilibrium spin density $\Delta s \propto \Delta V_s$. From Fig. 2d we see that with the respect to the $s_0$:
\begin{equation}
\begin{split}
\Delta s&=\frac{\mu_B}{e}(\Delta n^\uparrow - \Delta n^\downarrow) \\
&=\frac{\mu_B}{e}(-eV^\uparrow-s_0)\rho^\uparrow-\frac{\mu_B}{e}(-eV^\downarrow-s_0)\rho^\downarrow \\
&=-2\mu_B\Delta V_s\rho^\uparrow\rho^\downarrow/\rho e.
\label{eq11}
\end{split}
\end{equation}
Taking time derivatives we get from (9) and (10) the source term $f_s$ to be:
\begin{equation}
f_s=\frac{4I\rho^\uparrow\rho^\downarrow}{\rho}\frac{\Delta\rho}{\rho+4 I \rho^\uparrow \rho^\downarrow}\frac{\mu_B}{e}\frac{\partial n}{\partial t}=\gamma_s \frac{\mu_B}{e} \frac{\partial n}{\partial t}.
\label{eq10}
\end{equation}

In equilibrium the final change of magnetization $\delta m=\mu_B(\delta n^\uparrow-\delta n^\downarrow)/e$ is a consequence of uniform potential shift of the bands and their relative exchange splitting (Fig. 2e).

Equations (1)-(7) with source term (11) are solved numerically by finite element method in Comsol Multiphysics with time-varying voltage applied to the electrodes with amplitude $U_0=8$ V and frequency $f_0=200$ MHz. The relative permittivity is $\epsilon_r=10$ for MgO and $\epsilon_r=1$ for metals. We assume the conductivity of metals $\sigma=1.2 \cdot 10^7$ S/m and the diffusion constant $D=4 \cdot 10^{-3}$  m$^2$/s. The spin asymmetry coefficient for Co is $\beta=0.5$ \cite{Zhu2008}, the spin-flip relaxation time is $T_1=0.9$ ps \cite{Zhu2008} and the magnetoelectric coefficient calculated from Eq. (11) and parameters given in Ref. \cite{Zhang1999} is $\gamma_s=-0.25$. We assume also the magnetization of Co film lies in-plane of the film. Below we analize the distribution of amplitudes of spin accumulation $s$ and spin current $J_s$ between MgO layers with the contributions from SC and SS. 

\begin{figure}
\includegraphics*{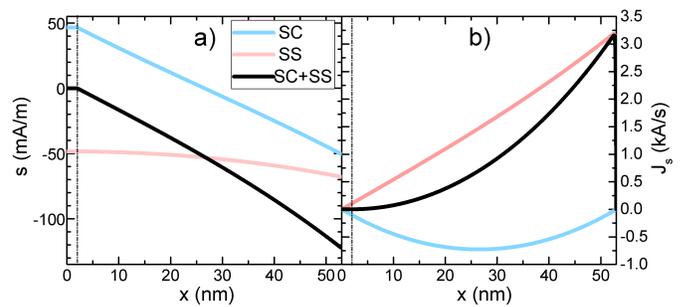}
 \caption{\label{Fig3} Spatial distribution of (a) spin accumulation and (b) spin current through the thickness of Cu-Co bilayer in system B with separated contributions from spin-dependent conductivity and spin-dependent surface screening. The dash-dotted line indicates Cu/Co interface.}
\end{figure}

Figure 3 shows the calculated spatial distribution of the spin accumulation and the spin current (for the system B and thickness of Co film $L_{\text{F}}$=51 nm) at the moment when they reach maximum values, i.e., when $V$ crosses zero value and charge current flows in the +$x$ direction. We separated the contributions which come from SC and SS. For the current driven by the spin-dependent conductivity, since $\beta>0$ then $\tau_- >\tau_+$, i.e., the relaxation time for spin down is bigger than for spin up and there is more spin down electrons flowing than spin up electrons. Consequently, $J_s$ has negative sign (Fig. 3b, blue line) and its maximum is at the center of Cu/Co bilayer. Within the distance proportional to the spin-diffusion length ($L_{sf}=$60 nm for Co) from the ferromagnet surfaces there builds up a negative spin accumulation gradient (Fig. 3a, blue line) that stands against SC term (compare Eq. (6)) \cite{Zayets2012}. In the Cu layer there is weak exponential decay of $s$ from the Cu-Co interface.

For the current driven by the spin-dependent surface screening, since $\gamma_s<0$, the outflow of negative charges from the F-D interface generates negative spin accumulation (Fig. 3a, red line). The resulting spin accumulation gradient induces spin current in the direction +$x$ (Fig. 3b, red line). The maximum value of $J_s$ is close to the D-FM interface.

We have shown that the spin current driven by SS is a consequence of the spin accumulation gradient build-up. This is in contrast to the SC-polarized current, where spin accumulation gradient is a result of this current and stands against it.

The effects from SS and SC sum up to give resultant spin accumulation and spin current (black lines in Fig. 3). In Co ($\beta>0$, $\gamma_s<0$) they add up destructively and for Co thickness of 51 nm discussed above, that contributions balance out at the Cu layer as shown at the left-hand side of dash-dotted line in Fig. 3b.

It is possible to distinguish between SC and SS generated spin currents comparing samples where SS would be absent (system A) and samples where SS would be present (system B). Moreover, spin-dependent conductivity is a phenomenon that acts in a bulk while spin-dependent surface screening is an interface effect. Thus, in thick ferromagnetic layer SC shall dominate while SS should manifest itself in thin ferromagnetic layers.

\begin{figure}
\includegraphics*{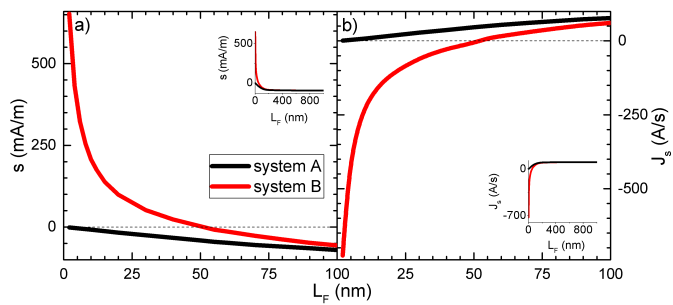}
 \caption{\label{Fig4} (a) Spin accumulation and (b) spin current in the Cu layer in dependence on the thickness of Co layer for system A and system B.}
\end{figure}

We consider the values of spin accumulation and spin current in the top Cu layer and we change the thickness of Co layer. The black lines in Fig. 4 show $s$ and $J_s$ calculated for the system A. Since there is no D-FM interface, the spin polarization of the current is a result of spin-dependent conductivity. The flow of spin-polarized charges builds up a spin accumulation as discussed above. For a thin Co layer ($L_{\text{F}}<2L_{\text{sf}}$) the spin accumulation counteracts the spin current significantly. For the thickness larger than $2L_{\text{sf}}$ the influence of the interfacial accumulation loses its significance and the value of spin-polarized current reaches maximum value of 105 A/s in Cu (and the value of $J_s  \approx -\beta J_f=7000$ A/s in the bulk of Co layer). The values of $s$ and $J_s$ saturate for thick Co films (see insets in Fig. 4).

The red lines in Fig. 4 show $s$ and $J_s$ for system B. Because of D-FM interface is present in the system, there are contributions both from SC and SS to the spin transport. Spin-dependent conductivity dominates for $L_{\text{F}}>L_{\text{sf}}$. If the distance from the D-FM interface to Cu layer is less than $2L_{\text{sf}}$, the currents induced by spin-dependent surface screening come into picture and the values of $s$ and $J_s$ deviates from those for system A. At $L_{\text{F}}=51$ nm SS balances out SC contribution in Cu and for $L_{\text{F}}<51$ nm spin current driven by SS dominates in the Cu layer. The maximum values obtained for $L_F=$ 2 nm are $s=$ 650 mA/m and $J_s=$730 A/s which are 7 times higher than those obtained for thick Co films by spin-dependent conductivity.

The dependences of $s$ and $J_s$ on $L_{\text{F}}$ in the systems considered above give us simple tool to distinguish between SS- and SC- driven currents. If SS contribution is absent in the system then $s \rightarrow 0$ and $J_s \rightarrow 0$ as $L_{\text{F}} \rightarrow 0$ for $L_{\text{F}}<L_{\text{sf}}$. If SS contribution is nonzero, $s$ and $J_s$ tend to nonzero values as  $L_{\text{F}} \rightarrow 0$ for $L_{\text{F}}<L_{\text{sf}}$. It is rather crucial to choose ferromagnet with relatively long spin-diffusion length like Co, since the effect is hard to resolve for $L_{\text{F}}>L_{\text{sf}}$.

In summary, the mechanism of spin current generation by the spin-dependent surface screening effect has been described within the Stoner model. The mechanism has been implemented to the drift-diffusion model. Then the specific dielectric-ferromagnetic heterostructures have been considered numerically to show the properties of SS-driven spin currents. We found the spin accumulation and the spin polarization of the current driven by spin-dependent surface screening is 7 times higher for 2-nm-thick Co than those obtained by spin-dependent conductivity for thick Co layers. Since the strength of the effect is dependent on the density of the screening charge at the D-FM interface, it may be enhanced with the use of dielectrics of high permittivity. Therefore, it is interesting to consider the influence of SS-driven spin currents onto magnetization dynamics, in particular, the possibility to decrease spin wave damping through currents induced by that mechanism. The proposed systems and the method of extraction of SS-driven spin accumulation is suitable to the X-Ray magnetic circular dichroism experiment \cite{Kukreja2015,Li2016}, where the measurement of spin accumulation dynamics is element-specific and thus may be constrained to the signal from Cu layer and directly compared to the dependences shown in Fig. 4.

\begin{acknowledgements}
The study has received financial support from the National Science Centre of Poland under grant 2018/28/C/ST3/00052.
\end{acknowledgements}

\end{document}